\documentclass{article}

\usepackage{arxiv}

\usepackage[utf8]{inputenc} 
\usepackage[T1]{fontenc}    
\usepackage{hyperref}       
\usepackage{url}            
\usepackage{booktabs}       
\usepackage{amsfonts}       
\usepackage{nicefrac}       
\usepackage{microtype}      
\usepackage{url}
\usepackage{hyperref}
\usepackage{fancyhdr}       
\usepackage{lipsum}
\usepackage{graphicx}
\usepackage[english]{babel}
\usepackage{subfig}
\usepackage{float}
\usepackage{amsmath,amssymb}
\usepackage{algorithm}
\usepackage{algorithmic}
\usepackage{bbm}
\usepackage{color}
\usepackage{biblatex}
\usepackage{csvsimple}

\def\hlinewd#1{%
\noalign{\ifnum0=`}\fi\hrule \@height #1 %
\futurelet\reserved@a\@xhline}

\usepackage{tabularx}
\usepackage{makecell}

\title{Efficient Hierarchical Storage Management Framework Empowered by Reinforcement Learning
}

\author{
  Tianru Zhang\\
  Department of Information Technology\\
  Uppsala university\\
  Uppsala, Sweden\\
  \texttt{tianru.zhang@it.uu.se}
  \And
  Salman Toor\\
  Department of Information Technology\\
  Uppsala University\\
  Uppsala, Sweden\\
  \texttt{salman.toor@it.uu.se}
  \And
  Andreas Hellander\\
  Department of Information Technology \\
  Uppsala university \\
  Uppsala, Sweden\\
  \texttt{andreas.hellander@it.uu.se} \\
}

\begin{document}

\maketitle

\begin{abstract}
With the rapid development of big data and cloud computing, data management has become increasingly challenging. 
Over the years, a number of frameworks for data management and storage with various characteristics and features have become available. Most of these are highly efficient, but ultimately create data silos. It becomes difficult to move and work coherently with data as new requirements emerge as no single framework can efficiently fulfill the data management needs of diverse applications. A possible solution is to design smart and efficient hierarchical (multi-tier) storage solutions. A hierarchical storage system (HSS) is a meta solution that consists of different storage frameworks organized as a jointly constructed large storage pool. It brings a number of benefits  including better utilization of the available storage, cost-efficiency, and use of different features provided by the underlying storage frameworks. In order to maximize the gains of hierarchical storage solutions, it is important that they include intelligent and autonomous mechanisms for data management grounded in the features of the different underlying frameworks. For example, it is essential that an HSS has a built-in data migration policy that determines the optimal placement of the datasets. These decisions should be made according to the characteristics of the dataset, tier status in a hierarchy, and access patterns. These are highly dynamic parameters and defining a policy based on the mentioned parameters is a non-trivial task. This paper presents an open-source hierarchical storage framework with a dynamic migration policy based on reinforcement learning (RL). We present a mathematical model, a software architecture, and an implementation based on both simulations and a live cloud-based environment. We compare the proposed RL-based strategy to a baseline of three rule-based policies, showing that the RL-based policy achieves significantly higher efficiency and optimal data distribution in different scenarios compared to the dynamic rule-based policies.

\end{abstract}


\section{Introduction}

The challenges of the data deluge  \cite{Gil2016, 6826494, 01270335} are traditionally summarized via the three Vs of big data: velocity, variety, and volume \cite{8258252}, which are  complemented by two additional V's - veracity and value \cite{7406335}. Big data challenges are not confined to a single discipline, rather they are equally valid for industry, academia, and individuals. According to a recent report published by Statista, “By $2022$, annual revenue from the global big data and business analytics market is expected to reach $274.3$ billion U.S. dollars” \cite{S1}. In terms of data, “In $2020$, it was projected that the overall amount of data created worldwide would reach $59$ zettabytes, climbing rapidly into the future” \cite{S2}. One of the most important factors in the data explosion is the growing IoT infrastructure that is considered to be vital for modern applications \cite{8961984}. When it comes to research and academia, the Large Hadron Collider \cite{LHC} experiment at CERN \cite{CERN} is a striking example where more than $30$ petabytes of data needs to be stored every year and 100 petabytes of data are already available for analysis \cite{LHCFF}. These are the numbers from a single research experiment and projects like human brain simulation \cite{HBP} and square kilometer array \cite{SKT} require far more resources to fulfill the data management needs of those projects. The scope and depth of the production of data does not stop here: according to a recently published report on DataPortal \cite{D1}, globally we have $4.2$ billion active users of social media. All these individual users are either generating new content or consuming already available datasets. 

The majority of available solutions for big data management are designed for specific needs. Often such solutions are limited in scope, usability, or data access requirements. However, the management of large datasets is a dynamic problem that cannot be solved with a single solution. The best strategy depends on the type of data, access patterns, usability of data, and long- and short-term availability of the data. To fulfill these requirements, we have centralized and distributed file systems that are optimized for different hardware (SSD and HDD) settings. There are tape storage solutions and also object stores for long-term storage. These solutions have different characteristics and are suitable for different storage requirements. However, it has been repeatedly reported that static availability of datasets cause issues related to performance and availability. Data access patterns change over time and there is no single storage solution that can address the dynamic requirements of varying access patterns. While there are some proprietary solutions based on hybrid data stores (volumes and object stores) that offer on-demand data transfer between different storage tiers, such solutions have limited capabilities and require significant resources. 

A recurrent challenge is the cost associated with the data stores. 
The cost of storage solutions depends on different factors such as access patterns, availability, security, and performance. As mentioned earlier, data access patterns change over time. To keep the data on the fastest storage solution (volumes based on SSDs) is economically challenging. On the contrary, pushing all the data on the slowest solution (object store) affects the overall performance of the applications. A possible middle ground solution is to store data on HDD volumes or distributing data between object store and HDD-based volumes. However, it is impractical to make these decisions manually; for example, how should one divide data between different storage frameworks? Or how long do data need to be stored in a particular storage solution? 

Another challenge that has become more prominent over the years is that not all the generated data are of high interest. Often different segments of the datasets become more interesting and there are more requests related to those segments. However, this process is also highly dynamic and interesting segments of the data vary over time or over different data analysis requirements. 

These are the core challenges that require careful thinking in order to address the needs of efficient and effective big data management solutions. Here it is important to note that the list of challenges also includes data security and privacy, as well as the development of new communication protocols and resilience. However, these are not the focus of the research presented in this article. 

To address the core challenges of intelligent transfer of data between different storage frameworks, cost-effectiveness, and access-aware dynamic placement of the data, we have designed and developed a high-level storage solution based on a hierarchical structure. The proposed solution consists of multiple storage frameworks. The underlying storage frameworks can be based on file systems using SSDs, HDDs, or even object storage. 

The key feature of the proposed solution is its capability of online, dynamic data transfer between different storage solutions, and to make decisions on the basis of data access patterns and usability. For this, we have used a Reinforcement Learning (RL) algorithm, which is the key enabler of dynamic placement of the data between different storage solutions. We have compared our proposed strategy with different rule-based policies, highlighting that while both reinforcement learning and rule-based approaches can achieve similar data distribution results based on dynamic data placement, the RL approach achieves this at a much lower cost compared to rule-based approaches. 

The main contributions of the article are as follows: 

\begin{itemize}
    \item We  present a well-defined theoretical model the hierarchical storage solution. The mathematical model is influenced by the article \cite{hsmrl} where the theoretical model is presented. We have made substantial modifications to build a real framework based on the theoretical model.
    
    \item We have designed and developed an open-source framework in Python. The code is available at GitHub and the framework has been tested in cloud environment settings. 
    
    \item We have done extensive experiments based on both simulations and real-life settings using a cloud environment. For a realistic comparison, we have also developed three rule-based schemes to compare with our proposed solution. 
    
\end{itemize}
 
The rest of the article is organized as follows: Section \ref{rlt} presents related work in the field and highlights our contributions. Section \ref{HSS} introduces the concept of a hierarchical storage solution and provides a mathematical model for the different settings. Section \ref{rule-based} explains the three rule-based policies we have used to compare our proposed reinforcement learning policy. Section \ref{system architecture} discusses the architecture and deployment of the storage solution both within simulation and real-world cloud settings. Section \ref{Results and Discussions} presents the results and related discussions based on both simulations and cloud-based experiments. The conclusion and future directions are part of Section \ref{Conclusion and Future Directions}. 
\section{Related work}
\label{rlt}


There are a number of successful distributed storage platforms such as GFS from Google \cite{gfs}, Cassandra from Facebook \cite{cassandra}, Ceph \cite{ceph}, Hbase \cite{hbase}, etc. These frameworks allow users of physically distributed systems to share their data and resources by using a common file system. 
However, most of the distributed system presented above were designed for single-level, i.e. storage blocks are all on the same level with the same or similar setups and properties. Plenty of different type of storage devices are available, including caches, SSDs, HDDs, tapes, object stores, etc. These devices usually have very different properties in terms of speed and volume. Single-level systems are not able to manage these devices simultaneously and merge them together. Therefore, the concept of hierarchical storage system (HSS) was invented. Hierarchical storage systems was first used by IBM in 2003 \cite{san}, where they came up with the SAN (Storage Area Networks). Based on SAN, they built many storage tanks and file systems with properties such as shared heterogeneous file access, centralized management and enterprise-wide scalability. Since then,  research have been carried out in terms of adding hierarchies into distributed file systems. This work includes adding hierarchy and heterogeneity into widely-used platforms. For example, an optimization for HDFS with a storage hierarchy \cite{GUAN2019415}, a three level hierarchical architecture for time-series data storage \cite{VILLALOBOS2020103257}, and DUX - an application-attuned dynamic hierarchical data management system for data processing frameworks \cite{7515715}.

There is a large volume of work in data management in cloud environments, for an overview see \cite{6826494}. Since clouds are heterogeneous environments that have various forms of storage services, a hierarchical structure is involved in data managing in the cloud \cite{7845021}. Another characteristic of cloud computing is the things are changing continuously with a varying rate of change. This seriously increases the difficulty of managing data manually. Therefore, self-managing dynamic cloud storage management has appeared. In \cite{8078264}, they discussed storage management algorithms which are both static and dynamic. In \cite{Torres2018}, a hierarchical model-based strategy was proposed and a case study of a cloud storage service was presented.

In order to achieve dynamic self-managing cloud storage management, various techniques were applied. The first class of options is using policy-based architectures, these attempts include STEPS architecture for IBM's SAN file system \cite{1410732}, LRU-K (Least Recently Used) replacement algorithm \cite{lru-k}, LFU (Least Frequently Used) replacement algorithm \cite{XING20181}, data placement systems \cite{4354142} (including PhEDEx \cite{phedex} and LDR \cite{ldr}), and eviction policies including LRU, LRFU, LIRS and ARC \cite{7847097}. 

Another popular option is to use online updating learning algorithm, such as Reinforcement Learning (RL). Reinforcement learning is a general class of algorithms in the field of machine learning that aims at allowing an agent to learn how to make automated decisions in an environment\cite{vanOtterlo2012}. It is an online automatic control system that adjust in real time, updating itself according to changes in the environment. For this reason, RL fits well with cloud data management, and it has been used in multiple cases in cloud computing, including cloud storage management\cite{Noel2019}, resource allocation \cite{7979983}, traffic optimization \cite{yqy20} and load balancing \cite{10.1145/3230543.3230551}.

Nevertheless, the work reviewed above has either focused on an hierarchical structure with fixed strategies, or only aimed at the overall optimization of the system instead of the  balance inside each tier. In this paper we develop a strategy to solve the data migration problem within hierarchical storage systems by using reinforcement learning. We propose a reinforcement learning based policy and three rule-based policies to manage the file migration process. We also build two frameworks for both a simulation environment and for a cloud environment, and run experiments using different frameworks and policies to study the advantages and disadvantages of each policy.

\section{Hierarchical Storage System (HSS)}
\label{HSS}
\subsection{Storage Hierarchies}

Hierarchical Storage System (HSS), also known as multi-tier storage system \cite{mts}, is a storage solution that connects multiple storage tiers in a hierarchical structure. Each tier in the structure consists of a dedicated storage framework. The included frameworks in the hierarchical structure together span multiple features. The list of features in general includes size, speed, efficiency, cost, security, and resilience. This list further includes custom features specific to the underlying framework's components and hardware in use. The aim with HSS architecture is to manage the combination of  different frameworks in a systematic and coherent manner. 



\begin{figure}[!h]
\centering
\includegraphics[width=2.5in]{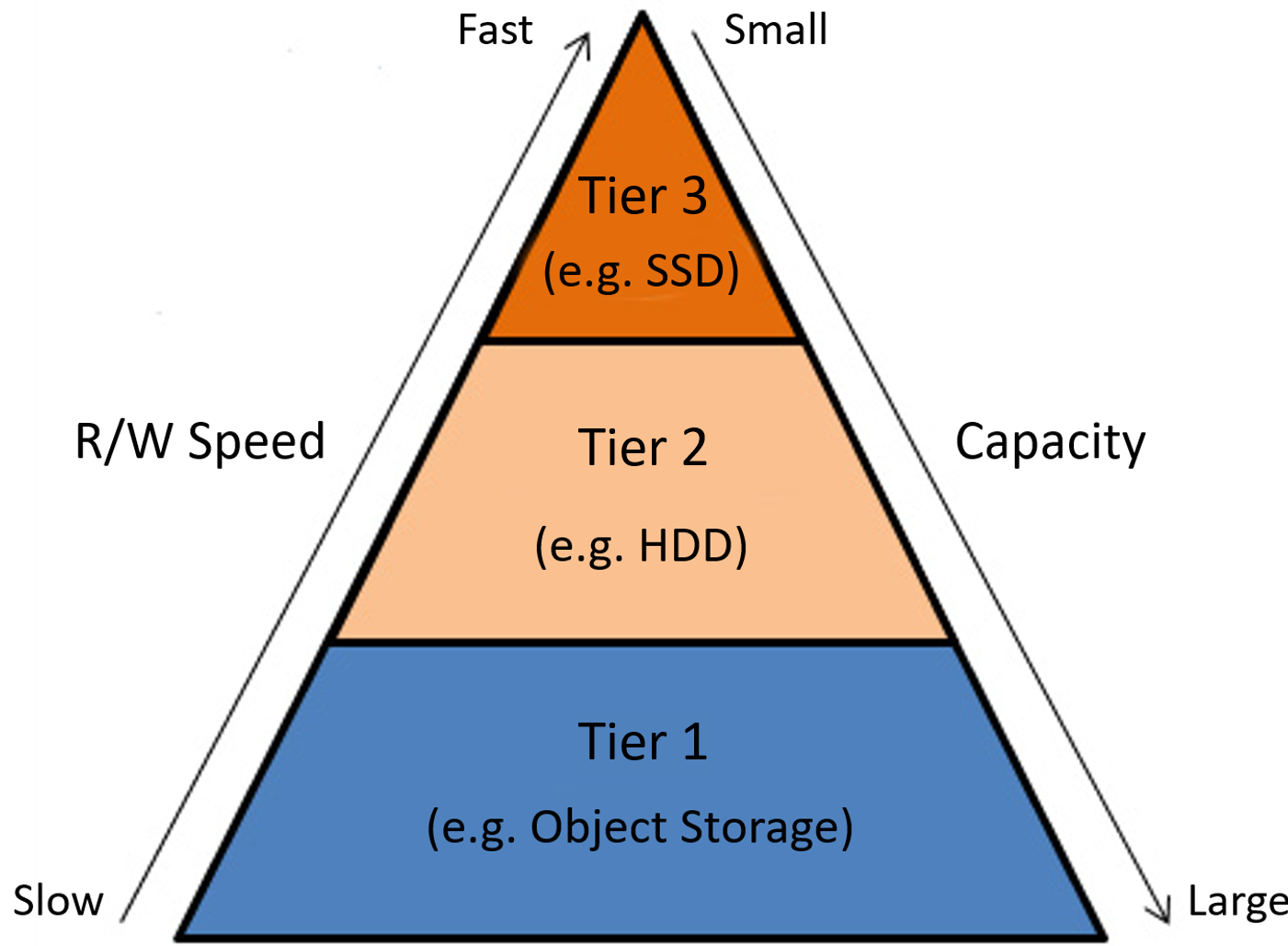}
\caption{Three tier hierarchical storage system.}
\label{fig_hss}
\end{figure}

Figure \ref{fig_hss} shows an example of a three-tier HSS. The hierarchy is defined by the access speed and storage capacity. Fast tiers are expensive and smaller in size. Whereas slow tiers are less expensive and significantly larger in size. This is mainly due to the internal architecture of the frameworks and underlying hardware (SSDs, HDDs or tape drives) used to build the solutions. 
A higher level or fast tier is used to store important and frequently requested datasets. Less important or less frequently accessed datasets are stored in slower tiers. The availability of datasets in an HSS can be stationary (available from an assigned tier) or dynamic(datasets automatically move between tiers). 
To realize the full benefits of the HSS, it should be equipped with autonomous, online and efficient data placement strategies to optimally utilize the available resources in different tiers.  

\subsection{Data migration policy}

A systematic way to decide which file should be stored in which tier is crucial. For example, if a file is being accessed frequently, then it should be stored in a fast tier to reduce the response time. Frequently accessed files are called \emph{hot} files, in contrast, files that are infrequently accessed are labeled as \emph{cold} files. The access frequency of a file is called \emph{hotness-level} and it is measured by the file temperature, which usually varies between $0$ and $1$. Where $0$ is the \emph{hotness-level} for a \emph{cold} file and $1$ for a \emph{hot} file.   

As the file properties (which in general include \emph{hotness-level}, size, type, and use-case dependent properties) change, files move between tiers. For instance, if one file is currently stored in fast tier, but it has not been requested for some time, then this file will be counted as a \emph{cold} file and moved to a slow tier in order to free up space for \emph{hot} files. 


The process of file migration consumes significant amount of compute and network resources. Thus a policy that can control these file migrations in an efficient way is important. Previous work on well-defined policies include Least Recently Used Replacement (LRU) \cite{lru-k}, Least Frequently Used Replacement (LFU) \cite{lfu}, Size-Temperature Replacement (STR) \cite{hsmrl}, and etc. However, these policies are mostly static and requires strict initial assumptions. However, data placement under varying access pattern is highly dynamic, and handling this becomes increasingly challenging with growing amount of the data in the system. To address the needs of autonomous data migration between tiers and efficient utilization of available resources, we have designed and developed an online data migration policy that adapts according to the incoming file access pattern, migrates data between tiers and consumes minimal system resources. The following subsection presents the mathematical model used to architect the online data migration policy.  



\subsection{Reinforcement learning-based policy}

Reinforcement learning (RL) \cite{RL} is a well-know approach for making decisions based on statistical estimations and environment variables. It aims at solving the Markov Decision Processes (MDPs). The MDP is an environment formed by $<S;A;P;R; 
\gamma>$, where $S$ is states, $A$ is actions, $P$ is transition probability, $R$ is rewards and $\gamma$ is the discount factor. The reinforcement learning aims at solving MDPs by learning the best policy($\pi:S\rightarrow A$) that gives the most correct action according to current state. The best action is decided so that the \emph{value function} and \emph{action-value function} under this policy and action are optimized. We refer to the class that executes these processes as an \emph{agent}. The agent will update its parameters when the system change from one state to another, and then perform the best action based on the current state and its parameters.

From above we can discover that the reinforcement learning is an online approach, since it updates its parameters simultaneously when target variable changes. This property of RL fits quite well with our demand of an online intelligent data migration policy. As long as we set the environment in the following way:
\begin{itemize}
    \item{states:} the property variables in each tier;
    \item{actions:} which file should be transferred to where;
    \item{rewards:} the cost signal (response time of HSS).
\end{itemize}
, we are able to define a MDP for a HSS and solve the migration policy by RL. 

This idea of using RL in hierarchical storage management has been discussed recent years, examples are mentioned in the related work section \ref{rlt}. Among all the researches, this paper \cite{hsmrl} presented a strong theoretical bases. Inspired by their work, we further explore the usage of reinforcement learning in hierarchical storage management.

In order to use RL in a HSS, the states variables are needed to be defined first. We use the following variables:
\begin{itemize}
    \item $s_1:$ average temperature of all files in the tier. This variable indicates the percentage of hot/cold files in the tier. A high value of $s_1$ indicates a large amount of requests, and thus a large expected cost of access.
    \item $s_2:$ average weighted temperature (temperature multiplied by file size) of all files in the tier. This variable presents further information about large/small files based on $s_1$. A higher value of this variable often implies larger files are relatively hotter, and thus will have a larger expected cost of access.
    \item $s_3:$ current queuing time in the tier. This variable denotes the queuing time for arriving requests. A high value implies a larger latency in this tier.
\end{itemize}

These states variables are all continuous variables in continuous time. However, the states of traditional MDPs are discrete and bounded in finite sets. Therefore, we are in the setting of an extended version of the MDP, the continuous time MDP, which also known as Semi-Markov Decision Process (SMDP) \cite{SMDP1}. 

The state-value function $v_{\pi}(s)$ of an MDP is the expected return starting from states $s$, following the policy $\pi$. In our case, we will use the cost function of each tier to represent the expected system response time starting from each states $s$. As discussed in the previous section, our problem is a SMDP whose states are continuous, therefore, to present the value function we need to use a function approximation. Various type of parameterized function approximation methods can be used to represent the value function, among them we choose the Fuzzy Rule-Based function (FRB) \cite{FRB}, for the reason that FRB function has relatively simple form and its parameters are easy to interpret. The rule-based function is a mapping $f$ from input vector $x\in \mathbb{R}^k$ to a scalar output $y$ by a combination of some rules. A common form of one rule is:
\begin{center}
    Rule $i$: IF $x_1\subset A_1^i, x_2\subset A_2^i,...,x_k\subset A_k^i$ THEN $p^i$
\end{center}
Where $x_1,...,x_K$ is the components of $x$, $A_1^i,...,A_k^i$ are fuzzy categories, and $p^i$ is the output parameter of this rule. The output of the rule-based function $f(x)$ is then a weighted average of $p^i$:
\begin{equation}
    y=f(x)=\frac{\sum_{i=1}^N p^iw^i(x)}{\sum_{i=1}^N w^i(x)}
\end{equation}
where $N$ is the number of rules, and $w^i(x)$ is the weight of rule $i$ computed by $w^i(x)=\prod_{j=1}^K\mu_{A_j^i}(x_j)$. $\mu_{A_j^i}(x_j)$ is the membership function, it takes values in $[0,1]$ and represent the measurement of an input variable $x_j$ belonging to category $A_j^i$, i.e. a value more close to 1 means $x_j$ highly probably belongs to $A_j^i$.

\begin{figure}[!h]
\centering
\includegraphics[width=2.5in]{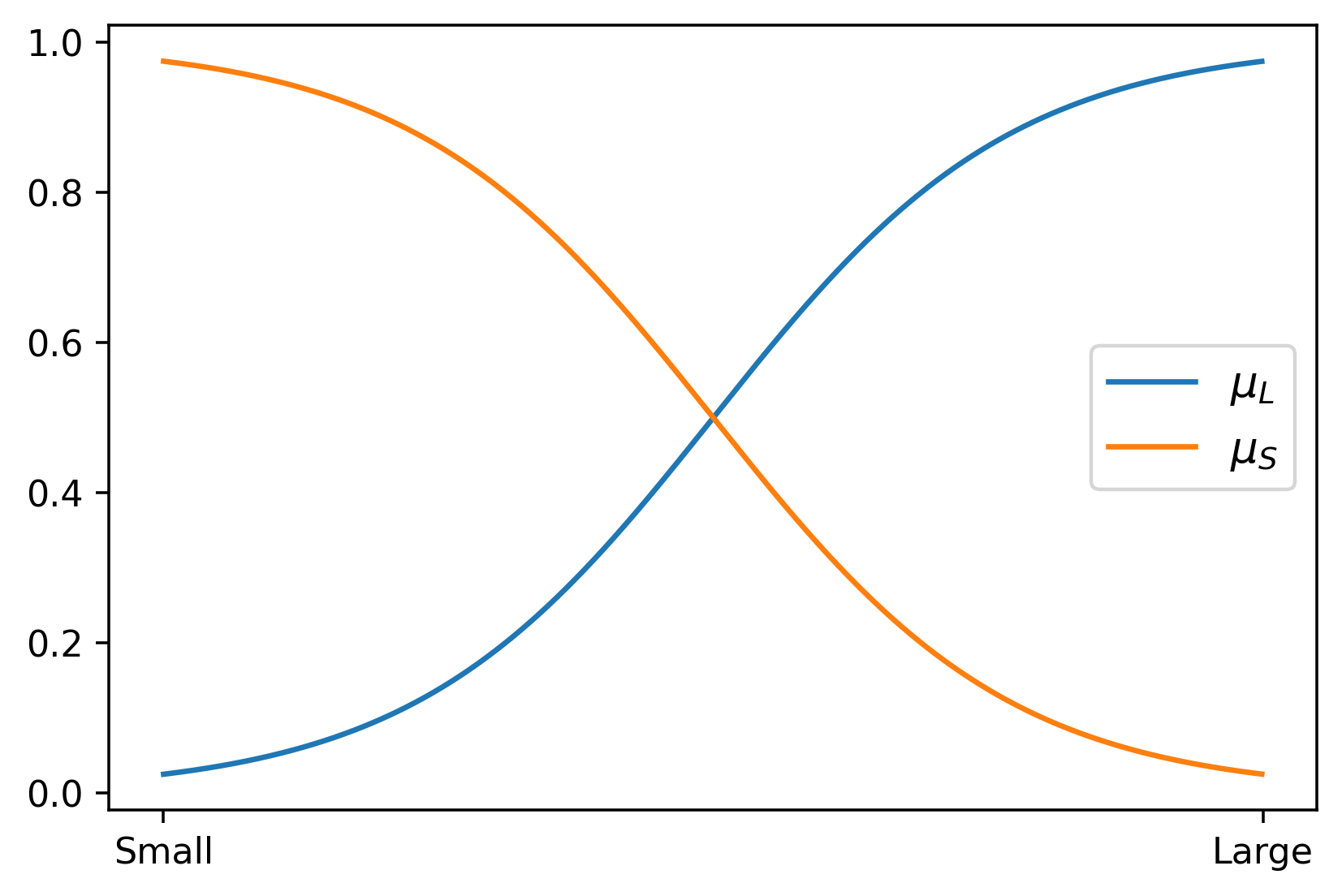}
\caption{Membership function representing the categories 'Large' and 'Small', the value of the function stands for how likely a input is 'Large' or 'Small'.}
\label{fig_memb}
\end{figure}

The FRB system is prevalent among modern industrial analysis, it usually contains large amount of complex rules to express a system. However, in our case the inputs are only the states variables, thus there is no need for a complex system. We choose a simple form that has only two categories $Small$ and $Large$ to describe the properties of inputs. And for the membership function, we choose a S-shape function $\mu_{Large}(x_j)=1/(1+a_je^{-b_jx})$, $\mu_{Small}(x_j)=1-\mu_{Large}(x_j)$. This kind of functions is suitable for cases that the ranges of $x_j$ are not constrained and no very large or very small values may occur. We can observe this property from the example figure \ref{fig_memb} of this pair of membership functions.

Based on the categories and membership function above, we formed a FRB function with $2^3=8$ rules as our value function.
\begin{equation}
    v(s)=\frac{\sum_{i=1}^8 p^iw^i(s)}{\sum_{i=1}^8 w^i(s)}
    \label{vfunc}
\end{equation}
where $s=(s_1,s_2,s_3)$ is the states variables, $w^i(s)=\mu_{S/L}(s_1)\mu_{S/L}(s_2)\mu_{S/L}(s_3)$ ($S/L$ represent $Small/Large$) is the multiplication of membership functions values at rule $i$.

After the states and value function are defined, actions need to be formed. For each file being requested, there are three situations. First, the file becomes hotter so that it will be upgraded to faster tier. Second, the file is not hot enough to be upgraded so that it will stay in its current tier. Third, there are other files which become hotter and upgraded from slower tier, but there is no enough space for the new hotter files in current tier and at the same time current file is the coldest file in current tier, then this file will be downgraded to slower tier in order to make room for new hotter files. For one timestep $t$, the action in this timestep is then formed by a set of all the files movement situation $\{file\ 1:up, file\ 2:down,...,file\ n:no\}$, where \emph{up} means upgrading to a higher tier, \emph{down} means downgrading to a lower tier, \emph{no} means no movement; and $file\ 1,...,file\ n$ are files being involved in timestep $t$.

The optimal placement of files in HSS requires each action to be taken carefully. It is important as there is a cost associated with each action.  
As we mentioned earlier, the policy is the key to take the actions. More precisely, it is used to control the file migrations between each tier. It works in the following way: When a R/W request is received for file $k$ (which is not already in the fastest tier), the policy decides whether this file should be upgraded to faster tier or not. In order to make such a decision between tiers $i$ and $j=i+1$, we define the migration policy in the following form: \emph{file k is upgraded from tier i to tier j if}
\begin{equation}
    \\
    C_{up}^i\cdot\Tilde{s_1}^i+C_{up}^j\cdot\Tilde{s_1}^j<C_{not}^i\cdot{s_1^i}+C_{not}^j\cdot{s_1}^j
    \label{ineq}
\end{equation}
where $C_{up}^{i/j}$ is the cost function of tier $i/j$ after file $k$ is upgraded, and $C_{not}^{i/j}$ is before upgrade; $\Tilde{s_1}$ is the average temperature of all files in tier if file k is upgraded, and $s_1$ is the same variable before upgrade. The tier average $s_1$ estimates the number of requests for files in the tier, and the cost function $C(s)$ predicts the future average requests response. Hence, their multiplication represents the average weighted response. Then the inequality (\ref{ineq}) denotes that the total response in the system is reduced, which means the overall efficiency of the system is increased.

The cost function should be updated as the file distributions in a tier changes. Therefore, in the following section we will discuss how to update the cost function. This kind of updating process is called 'control' in RL, and there are multiple control algorithms such as Monte-Carlo\cite{RL}, SARSA\cite{SARSA} and TD learning\cite{TD}. However, our application is a special case because we only have one fixed policy (\ref{ineq}) instead of multiple choices from a set of policies. For this reason, we use an off-policy control method to only optimize the value function called Temporal Difference (TD($\lambda$)) learning.

TD($\lambda$) is a well-known procedure for iteratively approximating the value function under a given policy. It updates the approximation of the value function (in our case the cost function $C(s)$) as follows:

\begin{equation}
\begin{aligned}
    \\\hat{C}^\pi(s)&=\hat{C}^\pi(s)+\alpha_n(R_n+\gamma\hat{C}^\pi(s_{n+1})-\hat{C}^\pi(s_n))z_n(s)\\\gamma&=e^{-\beta\tau_n}, z_n(s)=\lambda e^{-\beta\tau_n}z_{n-1}(s)+\mathbbm{1}(S_n = s)\\
    \label{td-c}
\end{aligned}
\end{equation}

where $\hat{C}^\pi(s)$ is the approximation of $C(s)$, $\alpha_n$ is the learning rate at the $n$th state, $R_n$ is the rewards until state $s_n$ (in our case we use a cost signal to represent to be introduced in (\ref{td-p}))),  $\gamma$ is the discounting factor (depending on the hyperparameter $\beta$), $\tau_n$ is the time the agent spends in state $s_n$, and $z_n$ is the eligibility trace, which is initialized to 0 and updated by the formula above and $\lambda$ is the trace-decay parameter of TD($\lambda$).

Recalling the formula (\ref{vfunc}), the cost function is actually a linear combination of basis functions with parameters $p^i$. Thus, we can rewrite the $C(s)$ in the form $C(s,p)=\sum_{i=1}^8 p^i\phi^i(s)$, where $\phi^i(s)=\frac{w^i(s)}{\sum_{i=1}^8 w^i(s)}$. Therefore, the updating of $C(s)$ (\ref{td-c}) using TD($\lambda$) amounts to updating the parameters $p^i$, so we can write the updating formula for $p^i$ in the following form:

\begin{equation}
\label{td-p}
\begin{split}
    p^i_{n+1}=p^i_n+\alpha_n(R_n+e^{-\beta\tau_n}&\hat{C}^\pi(s_{n+1})-\hat{C}^\pi(s_n))z_n^i(s) \\
    z_{n}^i(s)=\lambda e^{-\beta\tau_n}z_{n-1}^i&(s)+\phi^i(s_{n}) \\
    c_n=(1/X_n)\sum_{i=1}^{X_n} r_i&e^{-\beta(t_{n,i}-t_n)}
\end{split}
\end{equation}

where the eligibility trace $z_n^i$ is initialized to 0 and updated by \ref{td-p}. The rewards $R_n$ is given by the cost signal, where $X_n$ is the total number of requests in state $s_n$, $r_i$ is the response time of each request, $t_{n,i}$ is the arrival time of request $i$, and $t_i$ is the time arrived at state $s_n$. This iteration process is proved to converge for MDPs when the basis functions $\phi^i(s)$ are linearly independent \cite{TDFA}.

\begin{figure*}[!h]
\centering
\includegraphics[width=6in]{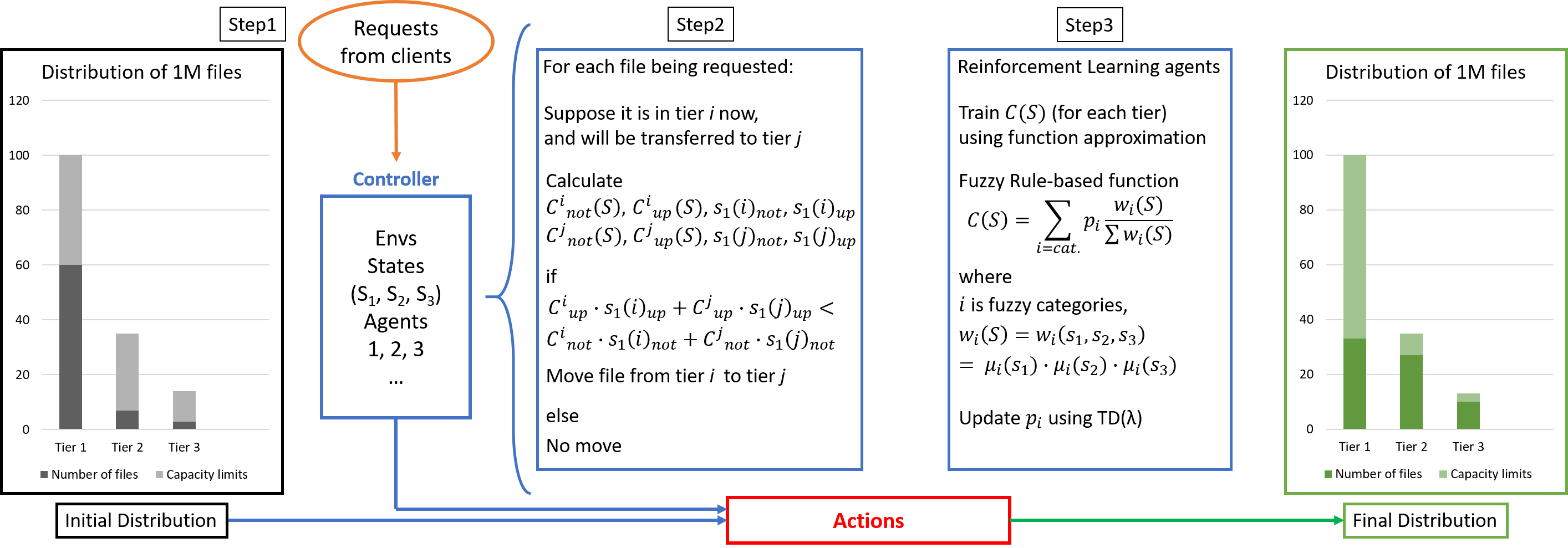}
\caption{Workflow of RL-based policy, formed on an example of three tiers HSS. Initial distribution of the files is not optimal (left side bar plot). After the RL-based policy controller takes actions based on the access pattern in step1 and the self-updating RL agents in step2 and step3, the framework optimally place files in different tiers.}
\label{flow_chart_rl}
\end{figure*}

The above mathematical definitions details the fundamentals of our RL-based policy for data migration in HSS. The algorithm \ref{alg100} presents a pseudo-code of the implementation. The figure \ref{flow_chart_rl} highlights the actual flow chart of the implemented framework based on a three tier hierarchical storage architecture. It includes actions, states and the three-step iterative approach to achieve optimal distribution of files in the HSS. Further details regarding the implementation and architecture are discussed in the section \ref{system architecture}.

\begin{algorithm}
\caption{RL-based migration policy}
\label{alg100}
\leftline{\textbf{Initial:} RL agents for each tier with initial parameter\{$a_i, b_i$\},} 
\leftline{\hspace{33pt}hyper-parameters of TD($\lambda$): $\lambda$, $\beta$}
\begin{algorithmic}
\STATE \emph{\#Requests split in timesteps}
\FOR{each timestep}
\FOR{each request}
\STATE Obtain $C_{up}$, $C_{not}$, $s_1$, $\tilde{s}_1$ from tiers and agents
\STATE Make migration decision by (\ref{ineq})
\ENDFOR
\STATE rewards = tiers.env(previous\_states, requests)
\STATE agents.update(rewards,previous\_paras)
\ENDFOR
\end{algorithmic}
\end{algorithm}

\section{Selected Rule-based Policies}
\label{rule-based}

As mentioned in the section \ref{rlt}, different efforts have been made to establish rule-based policies using LRU, LFU, and other algorithms. To compare our proposed online policy, we have devised policies that are based on common rules using file properties. More specifically, we defined three rule-based policies as following:

\begin{itemize}
    \item{\textbf{Rule-based 1.}} When new files come into the system, first store them in the fastest tier until reaching the $80\%$ of capacity limit. Then store the rest of the files in the second fastest tier, again until hitting the $80\%$ of capacity limit. Same strategy needs to apply to place all the files in rest of the tiers. The file migration process between tiers is based on the file temperature, hotter files will be upgraded and colder files will be downgraded.
    \item{\textbf{Rule-based 2.}} First store all files into the slower tier according to the tier's capacity. The file migration process is again based on the file temperature, hotter files will be upgraded and colder files will be downgraded.
    \item{\textbf{Rule-based 3.}} The initialization is same as in \emph{rule-based 1}. The file placement starts with the fastest tier. The file migration process is again based on file temperature. But the temperature change is in inverse ratio with the file size, i.e. a large \emph{cold} file needs more requests to become \emph{hot} than a smaller file.
\end{itemize}

These rule-based policies are direct and stationary policies that do not adapt with evolving data access patterns. The migration mechanisms are only based on the \emph{hotness level}. Therefore, these rule-based policies can have a good performance since they can keep the temperature hierarchy, i.e. hot files will be placed in fast tiers. We here consider these three policies as the baseline and compare their performance to the performance of our proposed highly dynamic RL-based policies. 

\section{System Architecture and Deployment}
\label{system architecture}

To test and evaluate the proposed RL-based and baseline rule-based policies, we have designed two environments: a simulation software and a cloud-based distributed framework for HSS. The simulation software is to test scenarios with different parameter settings, while the cloud framework is a real-world environment including complexities ranging from distributed infrastructure to heterogeneous system level dynamics. Here, we would like to highlight that it is a non-trivial task to build a framework for distributed infrastructure based on a comprehensive mathematical model and rigorously test the framework with different parameter settings. To ensure stability, correctness and accuracy of our RL-based distributed HSS framework, we equally value both the simulation software and cloud based framework. 

The open-source code repository of this project is published on Github \cite{github}. It includes code for both the simulation and cloud-based distributed framework. Our contribution based on simulation software helps our future research in the direction and can be beneficial for other practitioners to understand the policy dynamics before implementing new features for the real environment. 
\subsection{Simulation System}
\label{simuhsssec}

\begin{figure}[!h]
\centering
\includegraphics[width=4in]{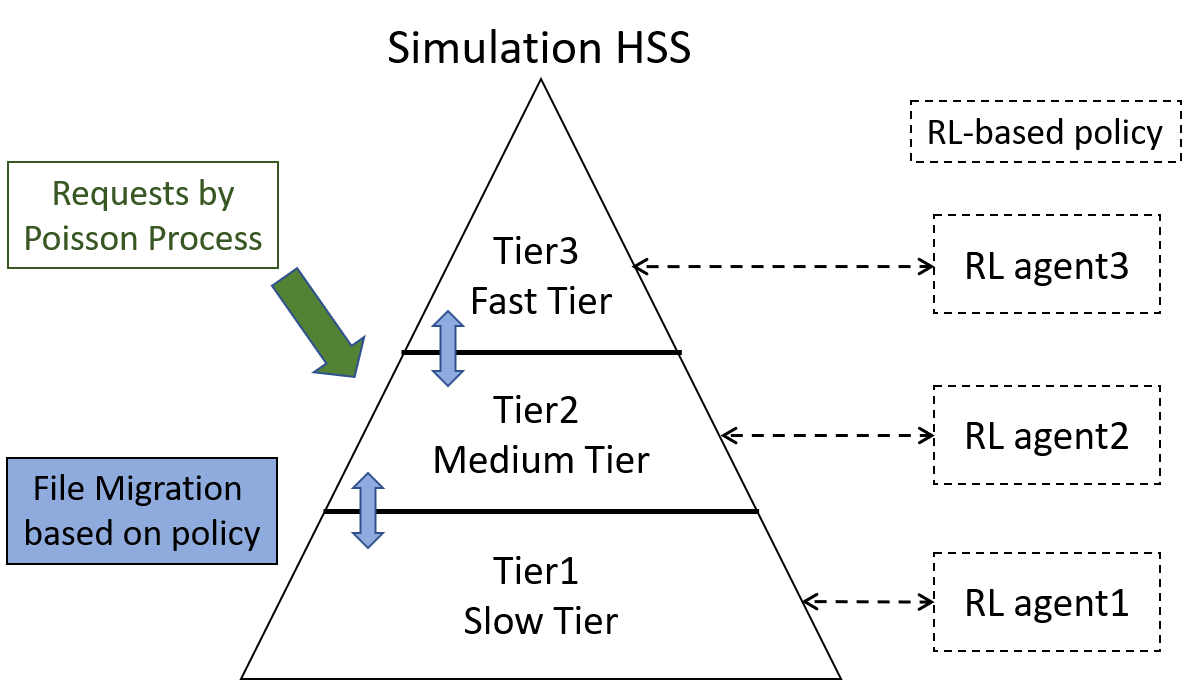}
\caption{Structure of the simulation system. Three tier hierarchical storage solution with varying access speed and storage capacities.}
\label{fig_simuhss}
\end{figure}

In the simulation software, we used a three-tier HSS structure. Figure \ref{fig_simuhss} highlights different components and the initial settings. To simulate the hierarchies in HSS, we design three tiers (Tier1, Tier2, Tier3) to represent tiers from slowest to fastest. Based on the relationship between access speed and storage capacity, a faster tier has a relatively smaller size. Therefore, we set a capacity limitation for Tier3 to be $100,000$, Tier2 to be $1,000,000$, and Tier1 to be $10,000,000$ units. Here it is important to note that in out experiments we assume that Tier1 (slowest tier) has large enough space to hold all the files.

We then created $1000$ files with size randomly distributed from $1$ to $10,000$, and temperature randomly between $0.1$ and $1.0$. The temperature of a file represent the \emph{hotness level}. The files with temperature higher than $0.5$ are \emph{hot} files, and files with temperature lower than $0.5$ are \emph{cold} files. For the $1000$ files, we have generated requests to simulate the access pattern. As discussed in previous literature, in real world the access (request) pattern of files in big data frameworks usually follows the Poisson distribution \cite{cao2001internet, TIAN2015135}. Therefore, we generate requests based on the Poisson arrival process. In order to ensure the read and write requests for each file, we have generated requests for \emph{hot} files using Poisson process with arrival rate of $0.5$, for \emph{cold} files using Poisson process with arrival rate of $0.01$. The request generating loop includes all $1000$ files at each timestep. With the distribution we set above, around $200$ requests were generated in each timestep. We have used $1000$ timesteps to run the simulations. 

The access requests change the \emph{hotness level} of files. For cold files, we set the probability to become hot after a request to be $0.3$. This probability is set to keep balance between hot and cold files in the system. For hot files, if they keep receiving requests then their temperature will not change, else if a hot file haven't received any request within $10$ timesteps, its temperature will be decreased by $0.1$. 



In the simulation software, when a file access request comes, the policy decides whether the file should be moved to another tier or not. The rule-based policies are only based on fixed rules, hence are deployed by adding the corresponding mechanism between each tier. For the RL policy, the decision is made by the criteria presented in the equation \ref{ineq}. It requires the calculation of cost functions, which are updated by the RL algorithm. Thus, we encapsulate the updating process and the decision process into an RL agent and attach one agent to each tier (right side of figure \ref{fig_simuhss}).

\subsection{Cloud-based distributed framework}
\label{cloudsyssec}
In this section we detail the architecture and deployment of our proposed HSS using cloud resources. In comparison with the simulation software, this implementation addresses the needs of a real world use case. Also this architecture is based on a three tier hierarchical structure. The three tiers are three storage platforms with different R/W speed and storage capacity. More specifically, we use three volumes with size of $2\ Gb$, $6\ Gb$, and $50\ Gb$; with R/W speed of $1000\ Mb/s$, $500\ Mb/s$, and $100\ Mb/s$. We have generated a dataset with $20,000$ files, with varying sizes between $10\ Kb$ and $200\ Mb$, and the total size of the whole dataset is $20\ Gb$.

\begin{figure}[!h]
\centering
\includegraphics[width=6in]{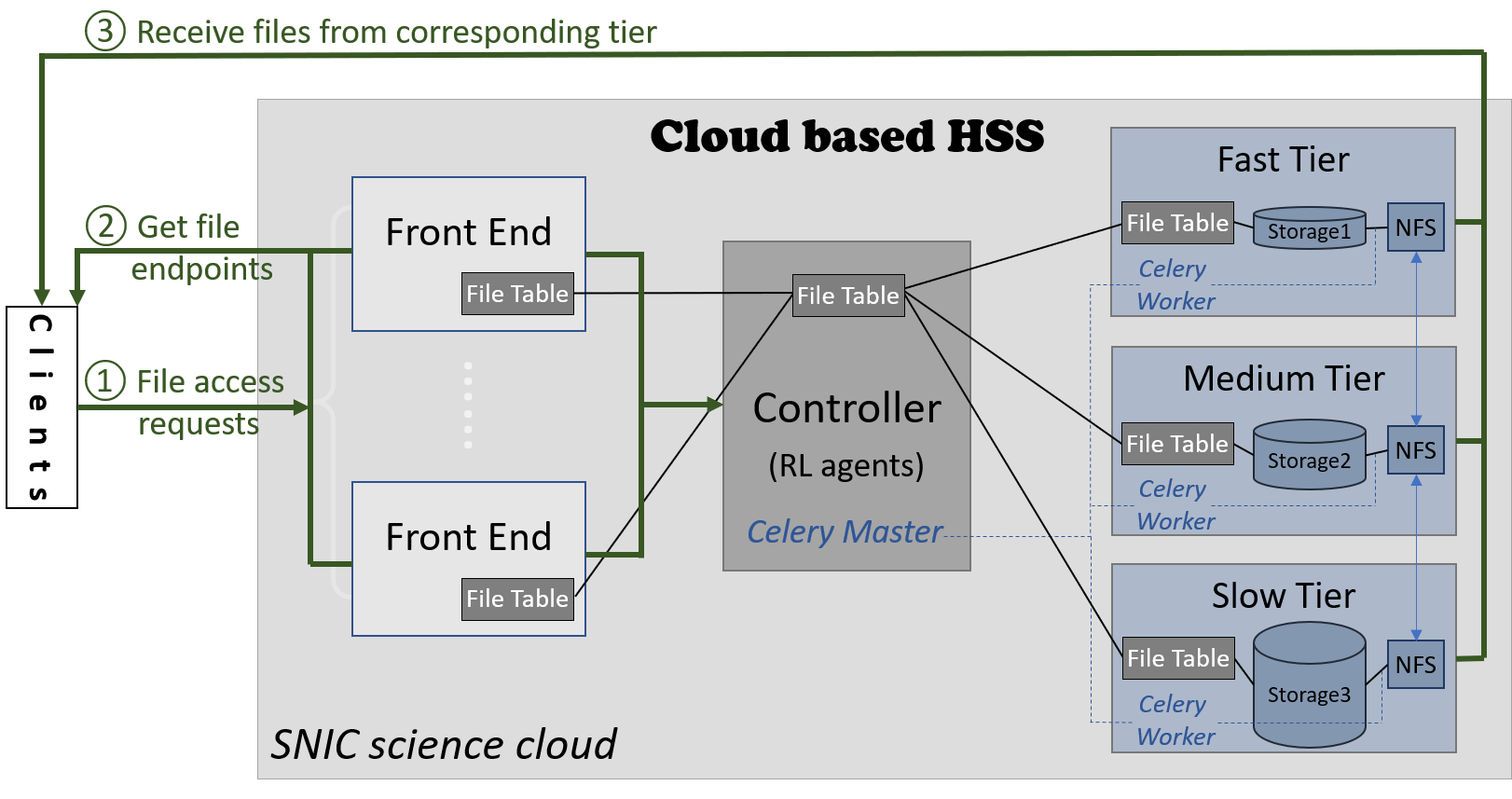}
\caption{Structure of the cloud distributed system. The architecture is based on three storage tiers, one controller node and multiple front-end nodes to handle requests coming from the clients.}
\label{fig_cloudhss}
\end{figure}

For this real-world deployment, we have used SNIC Science Cloud\cite{SSC}, a community cloud platform for Swedish researchers \cite{8109140}. 
The figure \ref{fig_cloudhss} highlights the framework components and communications between them. The three tiers are connected via RPC calls to be able to transfer files between tiers. 

Unlike the simulation setup, the three tiers in the cloud environment are not on one single machine. Thus to ensure a consistent system view between the RL agents, the controller and the storage tiers, we devised an metadata table that contains the status and basic properties of each file (size, temperature, access rights, file type etc.). 

Each tier in the hierarchy is responsible for its own metadata table. The information in different tables is gathered at the controller node. The controller node is a central component, responsible for executing the policies and taking migration decisions. 
In the case of rule-based policies, the controller outputs the decisions based on the given rules; as for the case of RL-based policy, the decisions are based on the equation (\ref{ineq}) and the RL agents for each tier. The decisions of migrating files are then sent to each storage tier by the messaging system. For the messaging system we use Celery \cite{celery}. We deployed the Celery master and RabbitMQ broker in the controller VM, and the Celery workers in the storage tiers. The transfers of actual files between tiers then happen separately. 

For the cloud HSS, we have implemented frontend nodes. The frontend nodes are connected with the controller to access the file information table. These frontend nodes are the entry points for the incoming requests. The nodes are responsible of receiving the requests from clients, update the controller about the incoming request, get the file location information from the locally available information table, generate a URL of the file and send it back to the client. As a final step, client access the file directly using the generated URL. Based on the defined interval or number of incoming requests, the controller initiate the migration decision as a background process. Here, it is important to note that the execution of the policies and decisions are completely independent from the process of data transfer. The execution of the policies is a background process that runs based on the defined intervals. 


The workflow of the cloud-based HSS framework can be described in the following steps: 

For the client side:
\begin{itemize}
    \item Send a read/write file request to the system (frontend node).
    \item Get file's endpoint(URL).
    \item Receive files directly from the corresponding tier.
\end{itemize}

For the controller and storage tiers, the framework perform the following steps as a background process:

\begin{itemize}
    \item Store new files into different storage tiers according to the initialization strategy of each policy;
    \item Record incoming requests, make file migration decision based on current file distribution, requests, and policy agents (only in RL case);
    \item Update policy agents (only in RL case);
    \item Transfer files between storage tiers according to the migration decision.
\end{itemize}

The steps in the controller and storage tiers are separate from the client's files access requests. For each client request, the file is directly transferred from the source (current tier location) to the destination(client). 

\section{Results and Discussions}
\label{Results and Discussions}

The presented results in this section include the following six file migration polices:  
\begin{itemize}
    \item{\textbf{Rule-based 1.}} Policy based on the first rule in sec.\ref{rule-based};
    \item{\textbf{Rule-based 2.}} Policy based on the second rule in sec.\ref{rule-based};
    \item{\textbf{Rule-based 3.}} Policy based on the third rule in sec.\ref{rule-based};
    \item{\textbf{RL-ft.}} RL-based policy with fast tier initialization. Decisions based on RL, with initially putting as many files as possible in the faster tier and the rest in the slow tier;
    \item{\textbf{RL-dt.}} RL-based policy with distributed tier initialization. Decisions based on RL, with initially putting 1\% of files in the fastest tier, 10\% of files in the medium tier, and the rest in the slow tier;
    \item{\textbf{RL-st.}} RL-based policy with slow tier initialization. Decisions based on RL, with initially putting all files in the slowest tier.
\end{itemize}

The first three provides the baseline rule-based polices. The next three are the proposed RL polices with three distinct initialization settings. We present results using both the simulation software and the cloud based real-world setup. 

To measure the effectiveness of the HSS we have used the \emph{hotness level} and file distribution in each tier. After each timestep, we recorded file temperature and distribution within each tier and plot this as a heatmap. Consider figure \ref{fig_heatmap1st} as an example. Each small block represents one file, and the color of the block indicates the file temperature. A darker color means a higher temperature. The heatmap also highlights the overall distribution of the files in the hierarchies. 

To measure the efficiency, we have checked the number of files being decided to transfer by each policy at each timestep. In the three tier structure, the transfer numbers we have taken into account are as following: 
\begin{itemize}
\label{num_transfer}
    \item Number of files being upgraded from Tier1 to Tier2
    \item Number of files being upgraded from Tier2 to Tier3
    \item Number of files being downgraded from Tier3 to Tier2
    \item Number of files being downgraded from Tier2 to Tier1
\end{itemize}

\subsection{Case-1: Simulation Results}
\label{case-1}
The presented results in this section are based on the simulation software (section \ref{simuhsssec}) using the mentioned six different policies. The detailed implementation can be found in \emph{simu\_experiments} folder of the Github repository \cite{github}. We have first generated $1000$ files as described in section \ref{simuhsssec}, then pushed them into three tiers according to the initialization criteria in each policy. Next, we built the policy between each tier. For rule-based policies, this entails creating migration mechanism. For RL-based polices, it entails creating RL agents with initial parameters and to set migration mechanism using RL agents. Further, we generated requests using the Poisson arrival process based on file temperatures, for $1000$ timesteps. We sent these requests into the system for each timestep and observe how the six policies perform in terms of the effectiveness and efficiency for the hierarchical storage system.

Based on the incoming file access requests, the temperature of the files started to increase. Meanwhile if a file haven't been requested for a long time, its temperature should be cooled down. Thus we define our mechanism of file temperature changing (named as \emph{hot-cold} function in implementation) in the following way: 
\begin{itemize}
    \item If a file is \emph{cold}, then after being requested it has a probability of 0.3 to become \emph{hot};
    \item If a file is already \emph{hot}, requests will not change the \emph{hotness level};
    \item If a file haven't been requested in $10$ timesteps, its temperature will decrease $0.1$ (until $0$).
\end{itemize}

The above mention criteria orchestrated the change in temperature. However, to start the process we still need to set initial temperature of each file. The initial temperature value can be set based on domain knowledge or any defined rule. For example, if we already know which files are more important, then those files can be assigned higher initial temperature values. While in the simulation case, since we randomly generate all the files, there is no pre-knowledge about the files. 
We have randomly assigned initial temperatures values between $0.4$ to $0.6$. This simple initial criteria allows files to easily switched between \emph{cold} and \emph{hot} at the early timesteps, which means the impact of the initial temperature is not too high.

\begin{figure}[!h]
\centering
\includegraphics[width=5in]{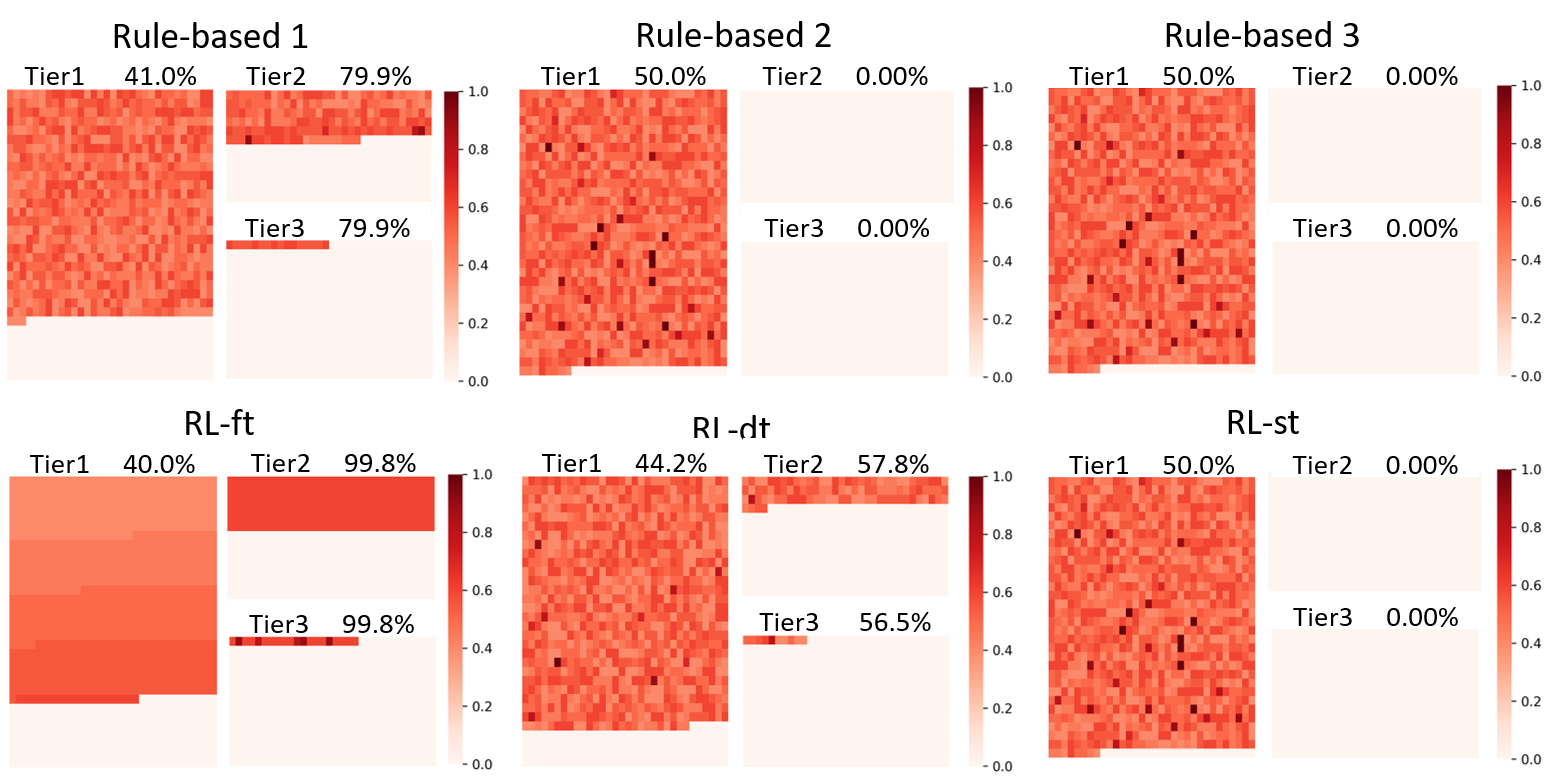}
\caption{File temperature distribution on the first timestep using each policy, illustrating the initialization of the tiers for each policy.}
\label{fig_heatmap1st}
\end{figure}

After setting up the policies, the framework ran for $1000$ timesteps. As discussed earlier in the section \ref{Results and Discussions}, we have used the heatmap and the number of transfers to measure the effectiveness and efficiency of using polices. The figure \ref{fig_heatmap1st} showcase the heatmap at the first timestep. Together with the file distribution between tiers, the figure also highlights the difference between the initialization methods of each policy and the storage capacity used in each tier (in percentage).

\begin{figure}[!h]
\centering
\includegraphics[width=5in]{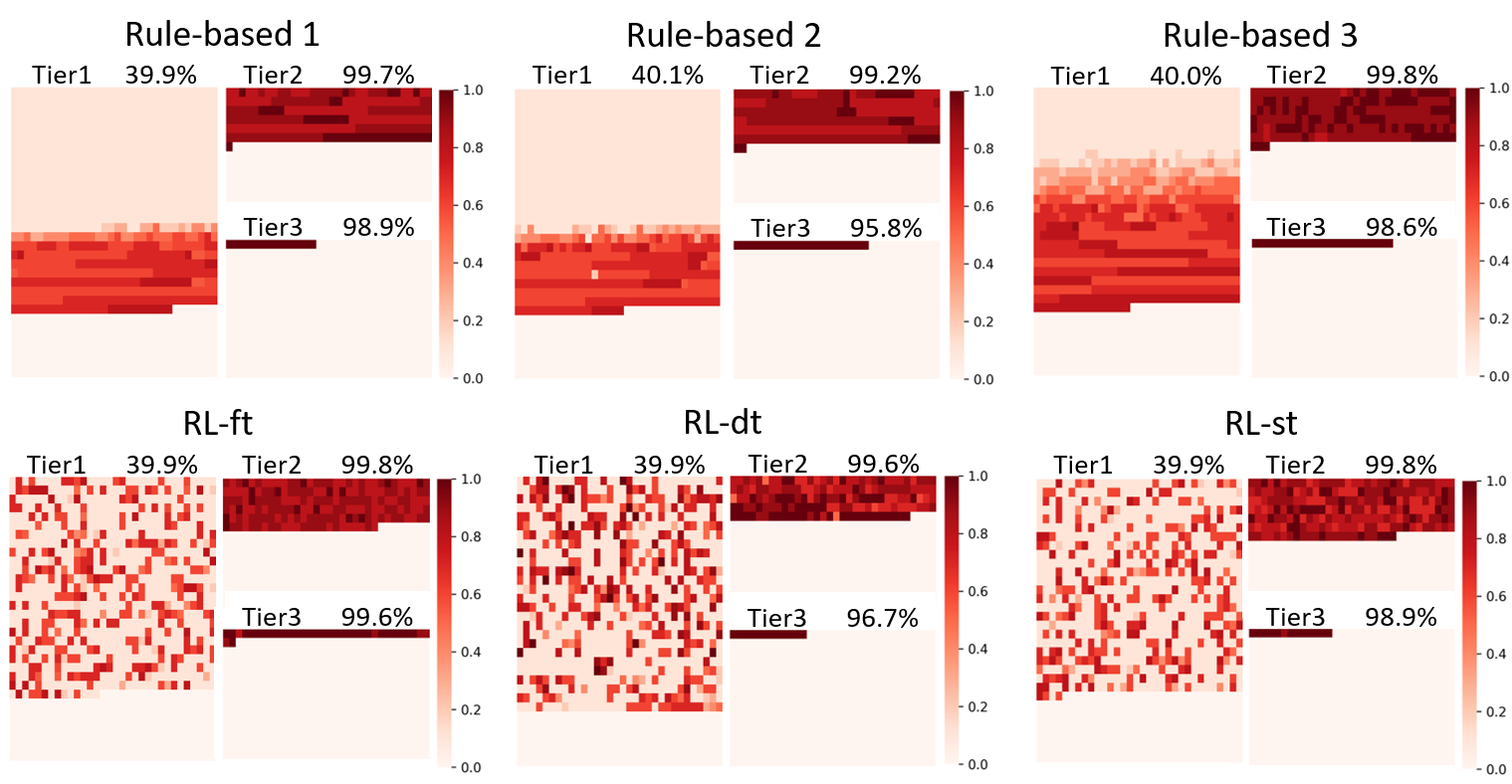}
\caption{Heatmap of files temperature distribution on the final timestep (1000th) using each policy.}
\label{fig_heatmap1000th}
\end{figure}

We ran experiments for $1000$ timesteps letting the files distribution in each tier be optimized according to the temperature. The figure \ref{fig_heatmap1000th} shows the heatmap at $1000th$ timestep (the final stage). The plot indicates that both rule-based and RL-based policies achieve ideal final states, in which hotter files are stored in a faster tier. The entire heatmap changing process can be seen as animation in the github repositpry \cite{github}. 

Another important finding is the optimal use of the available storage in each tier. Both Tier3 and Tier2 are more than $99\%$ full, and this final state is independent of the initial settings. The color of each file in Tier3 and Tier2 shows that all the important or frequently accessed files are optimally placed in the faster tiers. Here it is important to note that the distribution of the files between tiers is a continuous process based on the incoming requests. While the faster tiers are running at maximum capacity, if a file in Tier2 becomes hotter compare to some files in Tier3, the framework pulls the relatively less hot files from Tier3 to Tier2 to make room for the new \emph{hot} files. Thus, while the usage of the maximum capacity of each tier remains the same, the contents change based on the \emph{hotness level}.

Here it is important to note that the framework triggers the file transfer decision based on the equation \ref{ineq}. In brief, the file transfer decision only happens in the case of difference in \emph{hotness level}. The files with the same \emph{hotness levels} in different tiers do not trigger a transfer. This can be seen in the Tier1's status presented in the figure \ref{fig_heatmap1000th}. Tier1 is the slower tier but still contains relatively \emph{hot} files because the files in the faster tiers(Tier2, Tier3)  either have much hotter files or the temperature is the same as in Tier3.

To quantify the state in a numerical form, we have used the estimated response of the entire system. This variable represents the estimated future response of incoming requests under a given state. It is based on the following assumptions:
\begin{itemize}
    \item The request frequency of a file is positively correlated to its temperature, i.e. hotter file will have more requests
    \item The response time of a file is related to its size, i.e. a larger file will have a longer response time.
    \item The response time also effect by which tier a file is in, faster tier will lead to shorter response time. 
\end{itemize}

Based on the three variables above, we calculate the estimated system response for the six different policies. Their values are as follow:
\begin{table}[!h]
  \begin{center}
    \caption{Estimated system response under each policy.}
    \begin{tabular}{|c|c|c|c|c|c|} 
      \hline
      Rule-based 1 & Rule-based 2 & Rule-based 3 & RL-ft & RL-dt & RL-st\\
      \hline
      13480 & 13972 & 13931 & 13479 & 13591 & 13637\\
      \hline
    \end{tabular}
  \end{center}
\end{table}

The estimated system response also indicate the performance of rule-based and RL-based policies, that they are all highly effective in optimizing the file distribution in HSS.

However, the most significant finding is the utilization of resources to achieve the optimal distribution of files. For this, we observed the number of transfers (bullet points mentioned in the section \ref{num_transfer}). The difference between rule-based and RL-based policies is significant. We recorded the number of files being downgraded, and upgraded during each timestep and plot them. These numbers represent the efficiency of a policy, if one policy has a smaller transfer numbers  it triggers less file transfer processes, and eventually consume less resources (network, CPU and memory).

\begin{figure}[!h]
\centering
\includegraphics[width=6in]{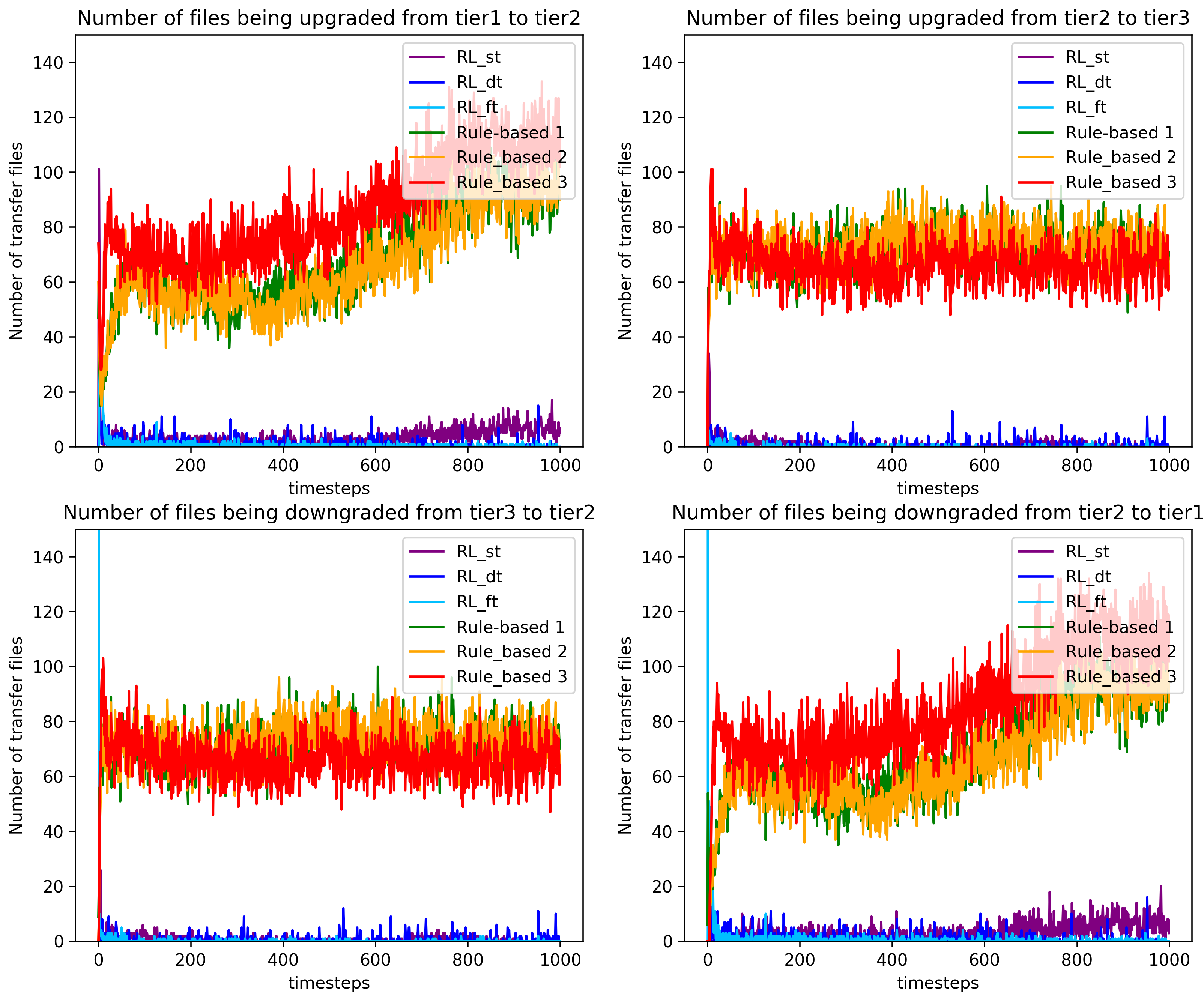}
\caption{Number of transfers between each tier under each policy.}
\label{fig_numtransfer}
\end{figure}

The figure \ref{fig_numtransfer} showcase the transfer numbers for rule-based policies and RL-based policies. The plot highlighted that rule-based policies lead to an much larger number of transfers (on average 80) per timestep. In comparison, RL-based policies required 20 (on average) transfers per time step. This indicates that the RL-policies outperformed the rule-based policies $5$ times in terms of efficiency. Although both rule-based policies and RL-based policies achieved ideal final stages, RL-based policies obviously cost much less resources to reach the same point.

In order to further strengthen our understanding with the results, we have performed more experiments with different scenarios to prove the consistency of RL-based policies. We have tested the policies with initial temperature of files between $0$ and $1$, instead of $0.4$ and $0.6$. The new temperature range is expected to create more chaos in the beginning states. We designed another experiment with the new temperature distribution. In the experiment, we have measured the number of transfers and estimated system response of the six policies to test their consistency.

\begin{figure}[!h]
\centering
\includegraphics[width=5in]{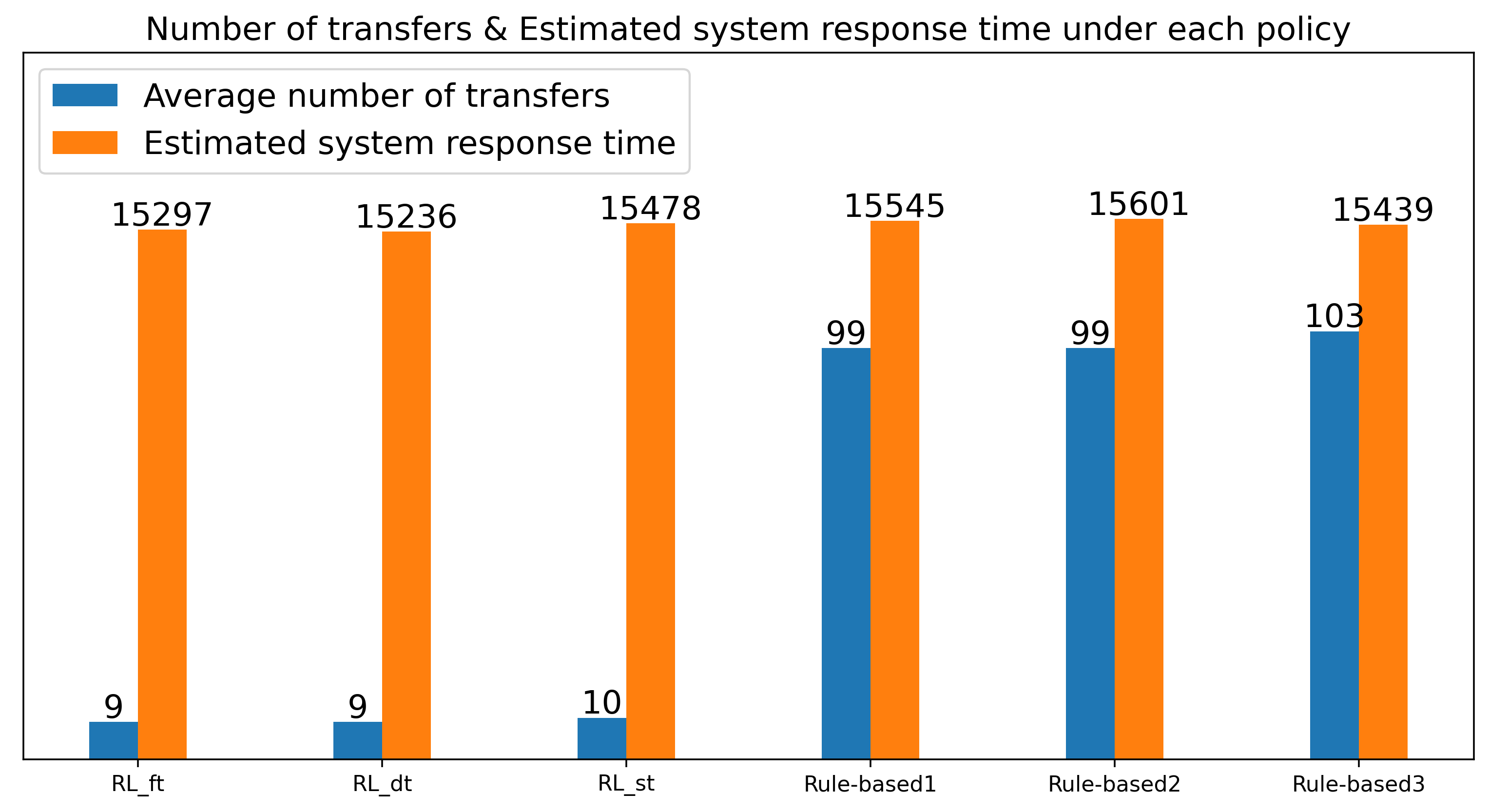}
\caption{Number of transfers and estimated system response under each policy, with initial temperature 0 to 1 and Poisson requests.}
\label{fig_numtrans&resp_temp01}
\end{figure}

The figure \ref{fig_numtrans&resp_temp01} presents the averages of number of transfers between tiers (average of the four numbers related to file downgrade and upgrade requests) and the estimated system response at the final timestep. We can see that the findings are similar to the previous experiment: the RL-based policies outperformed rule-based policies in terms of number of transfers. Meanwhile, both policies achieved a very close performance of file distributions at the final timestep.

The above experiments verified the consistency of RL-based policies under different initial temperature settings. However, the incoming access pattern is another factor that might effect the HSS and migration process. The above mentioned experiments are all based on access requests simulated by the Poisson arrival process (described in section \ref{simuhsssec}). Thus, it is important to test the policies with a different access pattern. 

To change the file access pattern, we have used the uniform distribution to select a fixed number of files to be requested in each timestep. More precisely, we have uniformly choose $200$ random files among the entire $1000$ files to be requested in each timestep. The requests of this pattern would give a higher chance for a \emph{cold} file to become \emph{hot}, hence there would be more \emph{hot} files in the end. 


\begin{figure}[!h]
\centering
\includegraphics[width=5in]{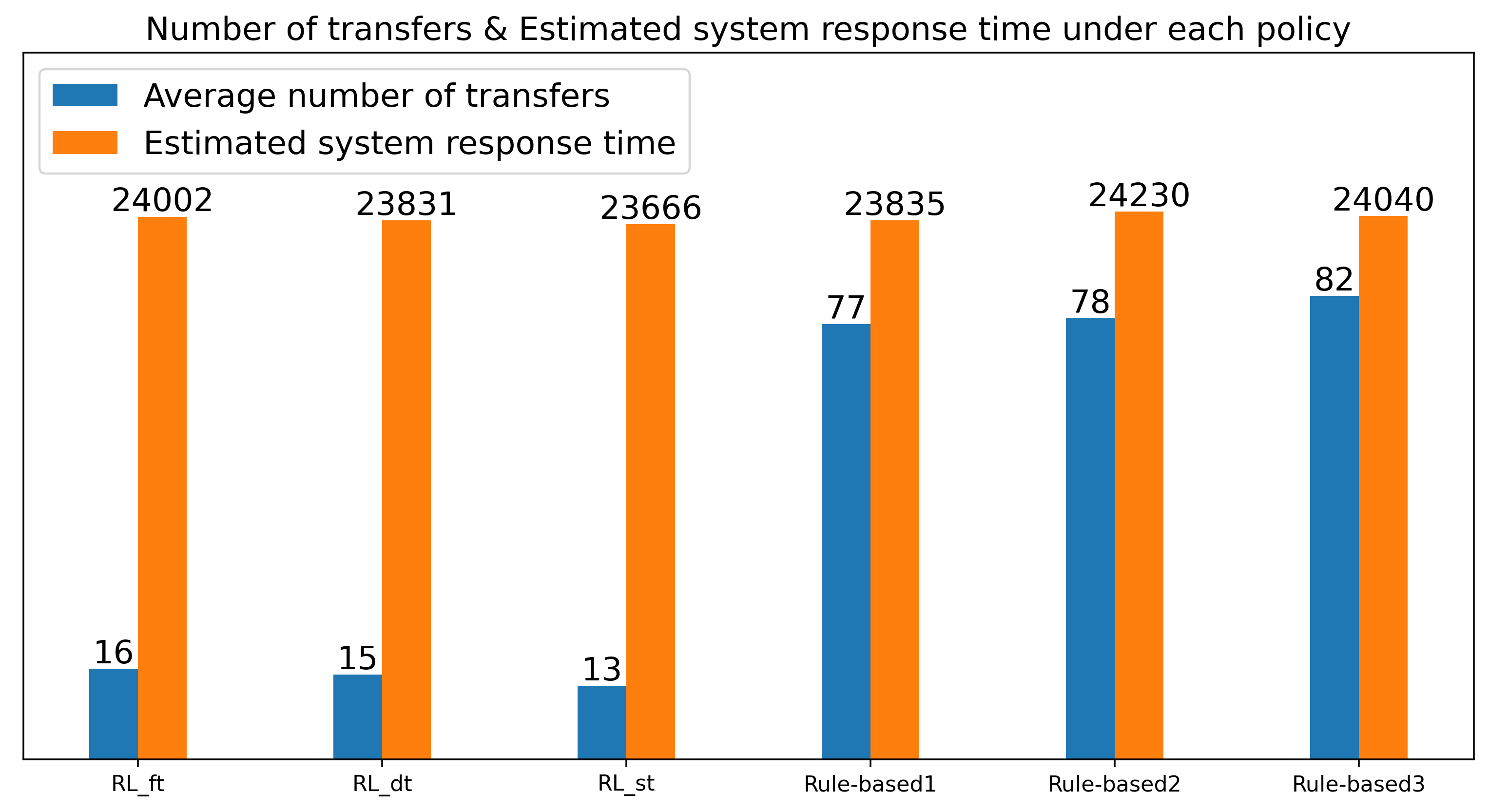}
\caption{Number of transfers and estimated system response time under each policy, with initial temperature 0.4 to 0.6 and uniformly random requests}
\label{fig_numtrans&resp_temp46_unifreq}
\end{figure}

The figure \ref{fig_numtrans&resp_temp46_unifreq} shows the results of using uniformly random requests. From the plot, we can conclude that the RL-based policies consistently perform well by achieving a similar optimal state as rule-based policies while creating much fewer file transfers. These results show the consistency of the RL-based policy under different request patterns.

\subsection{Case-2: Cloud Framework}
\label{case-2}

The experiments presented in the previous subsection were all based on the developed simulation software. The results clearly highlight that the effectiveness, efficiency and consistency of the RL-based policies are far superior than the rule-base policies. However, it is important to verify that the nice convergence of the RL-based file migration policies performs equally well while exposed to the heterogeneity of a large-scale real-world distributed environment. For this purpose, we have implemented and executed the RL-based policies and the rule-based policies in a real world cloud-based framework, introduced in section \ref{cloudsyssec}. The implementations were realized using code in directory \emph{cloud\_experiments} of the Github repository \cite{github}. The following subsections present our findings based on this real-world cloud setting.

\subsubsection{\textbf{Static Dataset Experiment}}
 
The first experiment is based on a static dataset with $20,000$ files, as described in section \ref{cloudsyssec}. We have generated $1\ million$ requests to these $20,000$ files. The incoming file access requests is a continuous process. Thus, we need to set a threshold to set the timestep interval so that the RL agents can update themselves after each timestep during the requests process. Since we have $1000$ timesteps in the simulation experiment and we observed good convergence of policies within $1000$ timesteps,  we set the threshold to update the file distribution to be $1,000,000/1000=1000$ requests, i.e. after $1000$ requests the RL agent updates its parameter.

For the temperature initialization, we have used a random value between $0.4$ to $0.6$. It provides a stable initial state by not having files that are too \emph{cold} or too \emph{hot}. Based on these settings, we then launched the cloud-based distributed framework. We initialized $20,000$ files in different tiers according to each policy and sent the $1 million$ requests. We used the same performance indicators as mentioned previously ( number of transfers and the estimated system response). 

\begin{figure}[!h]
\centering
\includegraphics[width=5in]{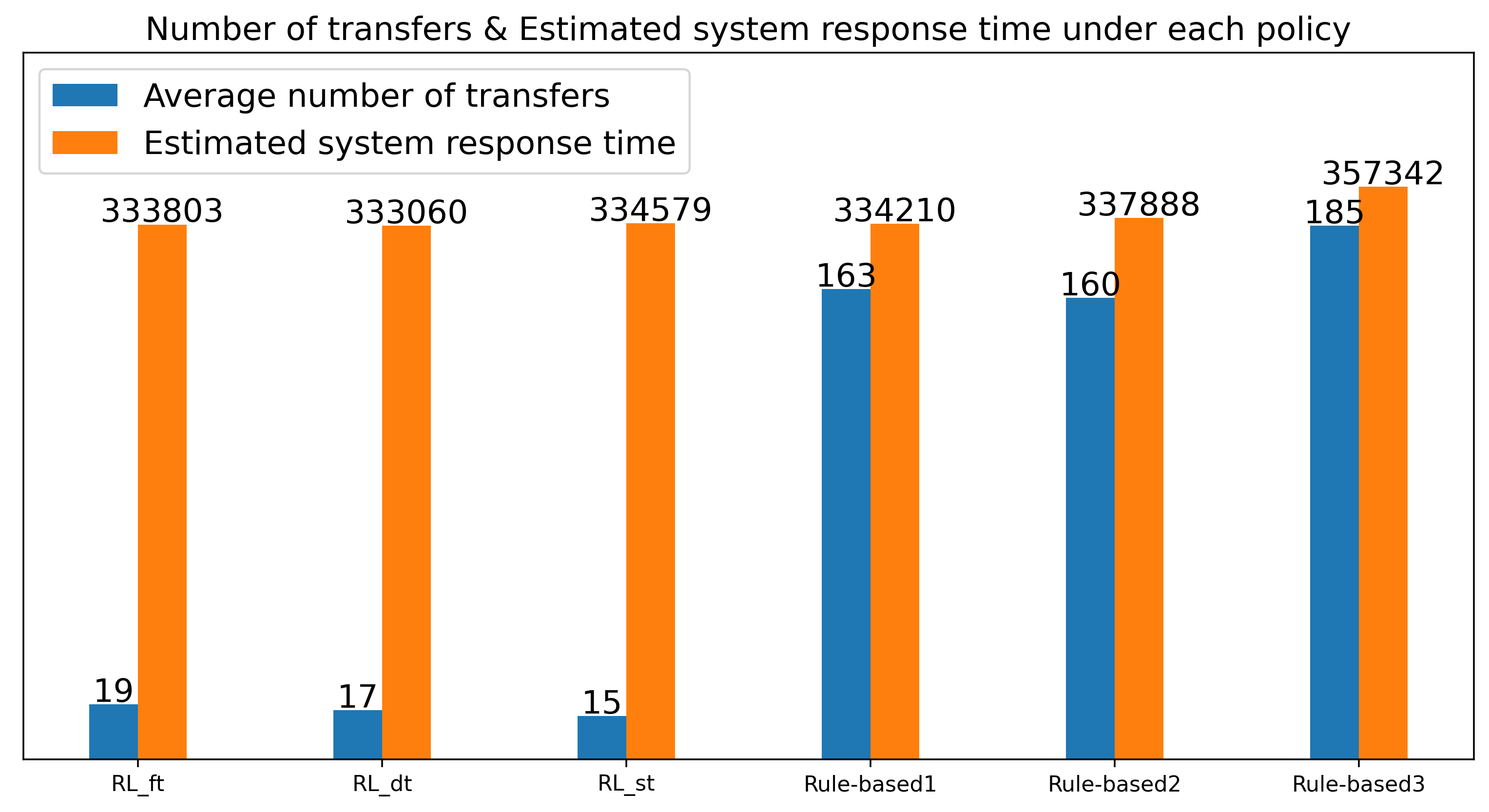}
\caption{Number of transfers and estimated system response time under each policy, cloud structure with initial temperature 0.4 to 0.6}
\label{fig_numtrans&resp_cloud_20000}
\end{figure}

The figure \ref{fig_numtrans&resp_cloud_20000} presents the result of the experiment described above, showing that there is no big changes in behavior between the cloud experiments and the simulation experiments, the RL-based policies are still more efficient, while achieving an approximately equal performance comparing with rule-based policies.

\subsubsection{\textbf{Dynamic Dataset Experiment}}
In the previous section, we discussed the case with a static dataset using cloud implementation. However, in th real world, storage systems are continuously receiving new data. Therefore, we next look at how the policies work in this setting. 

We designed another experiment starting from the final timestep ($1000th$ timestep) of the previous experiment. The settings are exactly the same, we used the request threshold $1000$ to update the policy, and run for $1000$ timesteps. In order to create the continuous workload, we pushed new files in the framework. First, we  generated $20,000$ new files with size between $10KB$ to $200MB$.  The $20,000$ new files were added as following: Add $200$ out of $20,000$ new files into the system after every 10 timesteps. The initial temperature of these files were between $0.4$ to $0.6$. Initially the new files were stored in the slowest tier because of the \emph{hotness level} and capacity limitation. 
We modified the access request pattern to be based on randomly selected files in the system. This request pattern allows newly arrived files to become \emph{hot} files.

\begin{figure}[!h]
\centering
\includegraphics[width=5in]{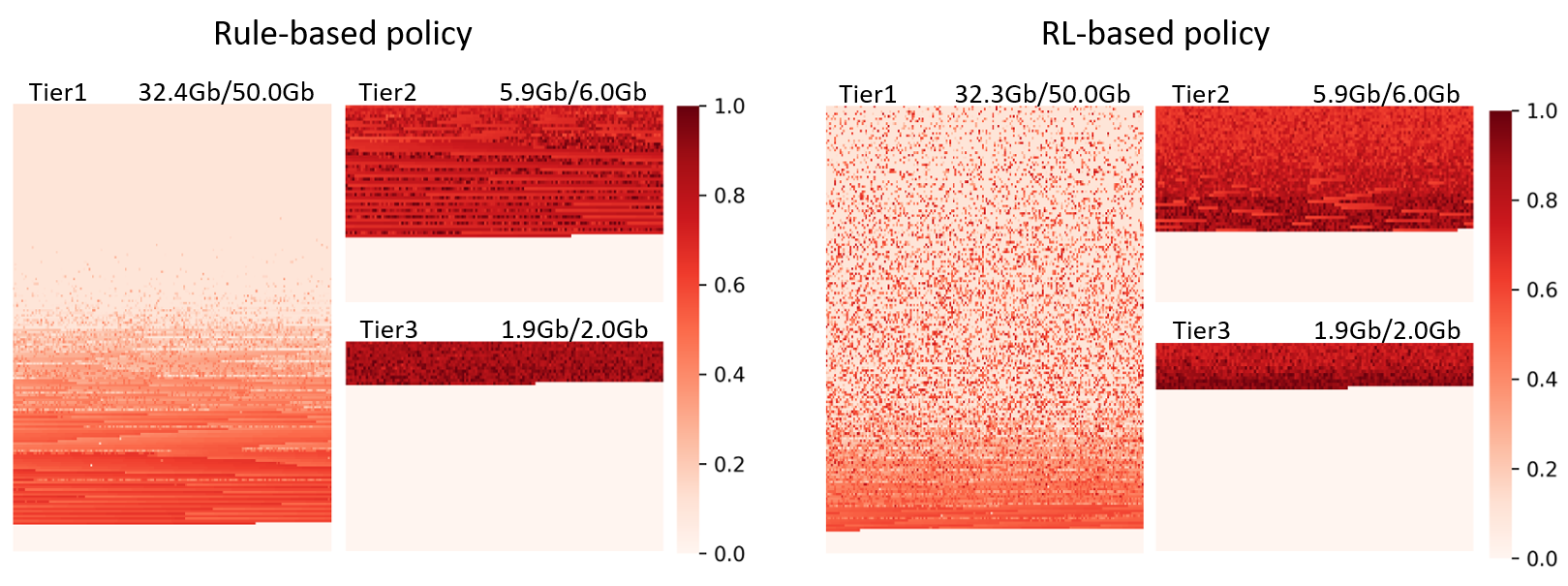}
\caption{File temperature distribution on the final timestep of cloud distributed system with dynamic dataset.}
\label{heatmap_20000+20000}
\end{figure}

\begin{figure}[!h]
\centering
\includegraphics[width=6in]{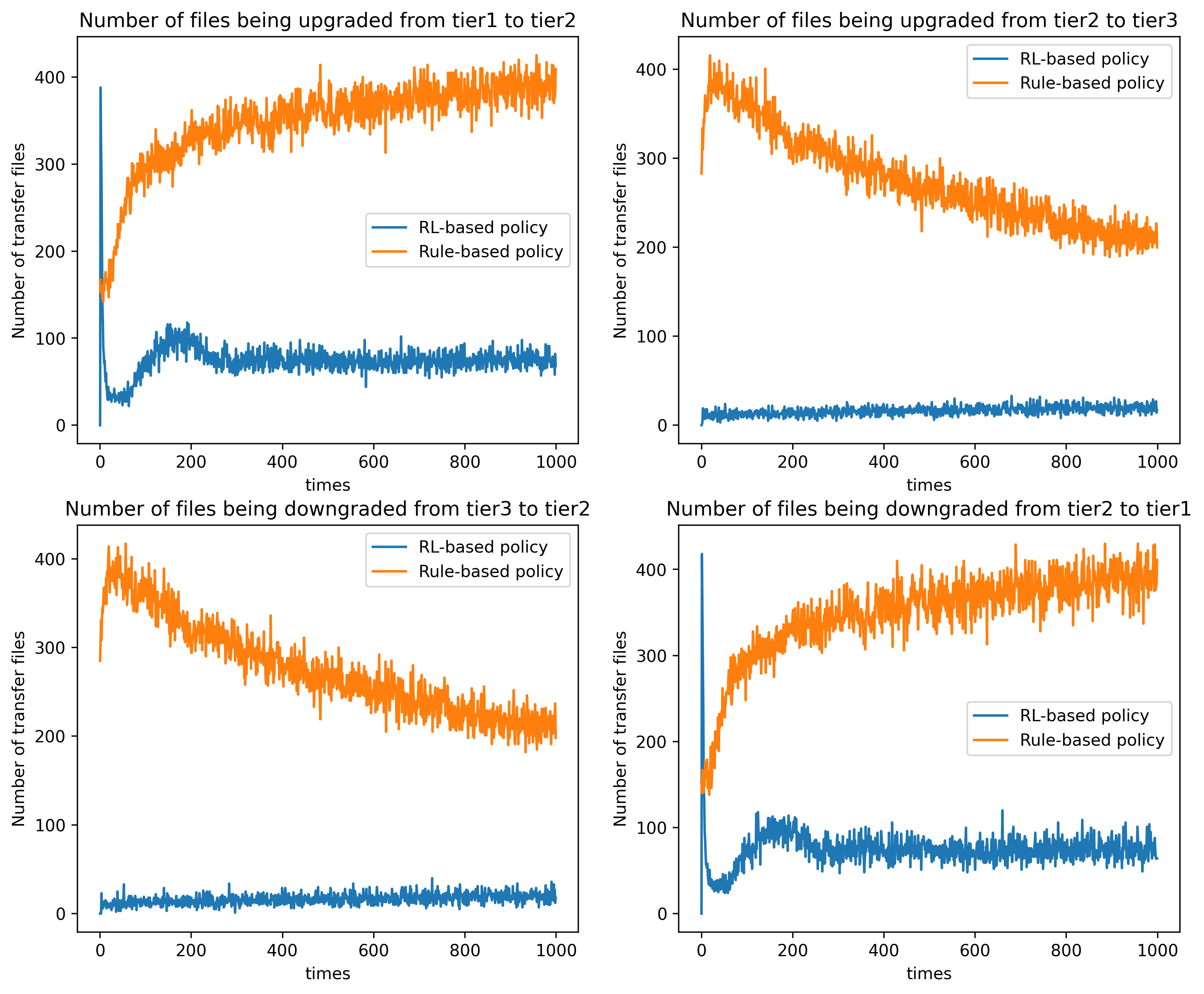}
\caption{Number of transfers between each tier based on the cloud distributed system with dynamic dataset.}
\label{fig_numtrans_20000+20000}
\end{figure}

Since we only have one initial file distribution criteria, we used one RL-based and one rule-based policy. The figures \ref{heatmap_20000+20000} and \ref{fig_numtrans_20000+20000} show the temperature distribution at the final state and the number of file transfers cost to reach the optimal state. 

Once again the results clearly highlight that the final states of the RL-based and rule-based policies followed the same pattern resulting in an optimal distribution of files --  the estimated system response (earlier discussed in the section \ref{case-1} ) was $696,977$ for the RL-based policy and $697,034$ for the rule-based policy.  Also the utilization of the storage capacity in different tiers is optimal, faster tiers hold hot files at maximum capacity. For more details on how the heatmap changes over time, see the animation in github repository \cite{github}.

The figure \ref{fig_numtrans_20000+20000} shows the required number of transfers. It is evident from the plots that the RL-based policy once again required fewer transfers than rule-based policy. The average numbers of transfers of the rule-based policy is $311$ at each timestep. While the RL-based policy required on average $45$ transfers per timestep. These results show that the RL-based policy is about $6$ times more efficient and effective than rule-based policy.


\subsection{Case-3: Performance and Stability}

The presented cases in subsections \ref{case-1} and \ref{case-2} have shown that both the RL-based and the rule-based policies can be used to obtain favorable file distribution in a HSS. While in terms of efficiency, effectiveness, and consistency, the RL-based policies outperformed the rule-based policies significantly (RL-based policies requires 5 to 6 times less transfers). The RL-based policy consists of its cost function, the RL agents and the properties of files in each tier (formula \ref{ineq}). All these components add additional complexity which requires more computations compared to rule-based policies. Hence, it is also important to consider the order of complexity and utilization of resources. 

Suppose there are $k$ timesteps and $n$ requests in each timestep. The RL-based policy does the calculation of equation (\ref{ineq}) for incoming requests and update the parameters in equation (\ref{td-p}) at each timestep. The computing complexity for the RL-based policy under each request pattern is $O(k\cdot n\cdot n_f)+O(k\cdot n_{para} \cdot n_f)$, where $n_f$ is the number of files in tiers and $n_{para}$ is the number of parameters $p^i$. For the rule-based policy, the computing complexity is $O(k\cdot n)$.

\begin{table}[!h]
  \begin{center}
    \caption{Execution time and memories consumption of each policy.}
    \label{resources}
    \begin{tabular}{|c|c|c|c|} 
      \hline
       Cases & \makecell[c]{Avg time per request} & memory consumption\\
      \hline
      rule-based simulation & 0.025 sec & 177 Mb\\
      \hline
      RL-based simulation & 0.040 sec & 178 Mb\\
      \hline
      rule-based cloud & 0.289 sec & 979 Mb\\
      \hline
      RL-based cloud & 0.521 sec & 984 Mb\\
      \hline
    \end{tabular}
  \end{center}
  
\end{table}

While running the experiments, we  recorded the execution time and the memory consumption of the policies. The table \ref{resources} presented the average execution time per decision and the memory utilization.

As can be seen, the RL-based policy requires slightly more execution time than the rule-based policy, it aligns well with our discussion regarding the computational complexity. As for the memory consumption, there is no big difference between the policies. This is mainly because that the memory consumption is mostly for the metadata information table of files and it is required by the each policy, thus the policies use basically the similar memory footprint.

\section{Conclusion} 
\label{Conclusion and Future Directions}

In this work we explored the possibility of using reinforcement learning in hierarchical storage management by using an RL algorithm to train a migration policy. We presented both the theoretical support of using RL in defining the data migration policy 
as well as designed and implemented a simulation software and a cloud-based framework to test the RL-based policy. The experiments based on these two HSSs were carried out to evaluate the performance of RL-based policies. The presented rule-based policies are designed and implemented as a baseline. Our experiments in terms of the  distribution of files and the number of transfers need to achive that optimal distribution proved the effectiveness, efficiency and consistency of the RL-based policy. For all the presented cases, the RL-based policy outperformed the rule-based policies. Despite the RL-based policy having a slightly higher computational complexity due to the usage of the RL algorithm,  the gain from less data transfer between tiers, favorable placement and adaption based on incoming data access patterns are significant benefits that cannot be achieved by other strategies. Therefore, we believe that our results show  that the use of an RL-based data migration holds great promise to gain maximum benefits from  large-scale hierarchical storage management solutions.


\section*{ACKNOWLEDGMENTS}

This presented research is supported by the Swedish Foundation for Strategic Research(SSF) under Grant No. $BD15-0008$. We would also like to acknowledge Swedish National Infrastructure for Computing (SNIC) for providing cloud resources, project number, $SNIC\ 2020/20-42$ and support from eSSence project, a strategic collaborative research programme in e-science.

\printbibliography

\end{document}